\documentclass[%
 reprint,
superscriptaddress,
 amsmath,amssymb,
 aps,
 prapplied 
]{revtex4-1}

\usepackage{graphicx,siunitx}% Include figure files
\usepackage{dcolumn}% Align table columns on decimal point
\usepackage{bm}% bold math
\DeclareSIUnit\Molar{\textsc{m}}

\begin{document}

\preprint{APS/123-QED}

\title{Photochemical Upconversion Theory:  Importance of Triplet Energy Levels and Triplet Quenching}% Force line breaks with \\
%\thanks{A footnote to the article title}%

\author{David Jefferies}
\affiliation{ARC Centre of Excellence in Exciton Science, School of Chemistry, Monash University, Clayton, Victoria, Australia}
\author{Timothy W. Schmidt}
\affiliation{ARC Centre of Excellence in Exciton Science, School of Chemistry, UNSW Sydney, Sydney, NSW, Australia}
\author{Laszlo Frazer}
\affiliation{ARC Centre of Excellence in Exciton Science, School of Chemistry, Monash University, Clayton, Victoria, Australia}

\date{\today}% It is always \today, today,
             %  but any date may be explicitly specified

\begin{abstract}
Photochemical upconversion is a promising way to boost the efficiency of solar cells using triplet exciton annihilation.  Currently, predicting the performance of photochemical upconversion devices is challenging.  We present an open source software package which takes experimental parameters as inputs and gives the figure of merit of an upconversion system, enabling theory-driven design of better solar energy devices.  We incorporate the statistical distribution of triplet excitons between the sensitizer and the emitter.  
Using the dynamic quenching effect of the sensitizer on emitter triplet excitons, we show that the optimal sensitizer concentration can be below the sensitizer solubility limit in liquid devices.  
These theoretical contributions can explain, without use of heavy atom-induced triplet exciton formation or phenyl group rotation, the experimental failure of zinc octaethylporphyrin to effectively sensitize diphenylanthracene, where platinum octaethylporphyrin succeeds.  
Our predictions indicate a change in direction for device design that will reduce triplet exciton losses.
\end{abstract}

%\pacs{Valid PACS appear here}% PACS, the Physics and Astronomy
                             % Classification Scheme.
%\keywords{Suggested keywords}%Use showkeys class option if keyword
                              %display desired
\maketitle

%\tableofcontents

\section{Introduction}
Solar cells have a transparent region below their bandgap.  The transparent region plays an important role in limiting the efficiency of conventional solar cells illuminated by sunlight \cite{shockley1961detailed,hirst2011fundamental}.  Photochemical upconversion is a phenomenon which converts light a solar cell cannot use into light that the cell can use \cite{2017recentpedrini,frazer2017optimizing,zeng2017molecular,schulze2015photochemical,mccusker2016materials,gray2018towards,chen2018upconversion}.  The utility comes from the spontaneous increase in the energy per photon.  Owing to its exothermic nature, photochemical upconversion can be relatively efficient \cite{wu2016solid,mccusker2016materials,frazerphotochemical,baldo2000transient}.

\begin{figure*}
    \centering
    \includegraphics[width=\textwidth]{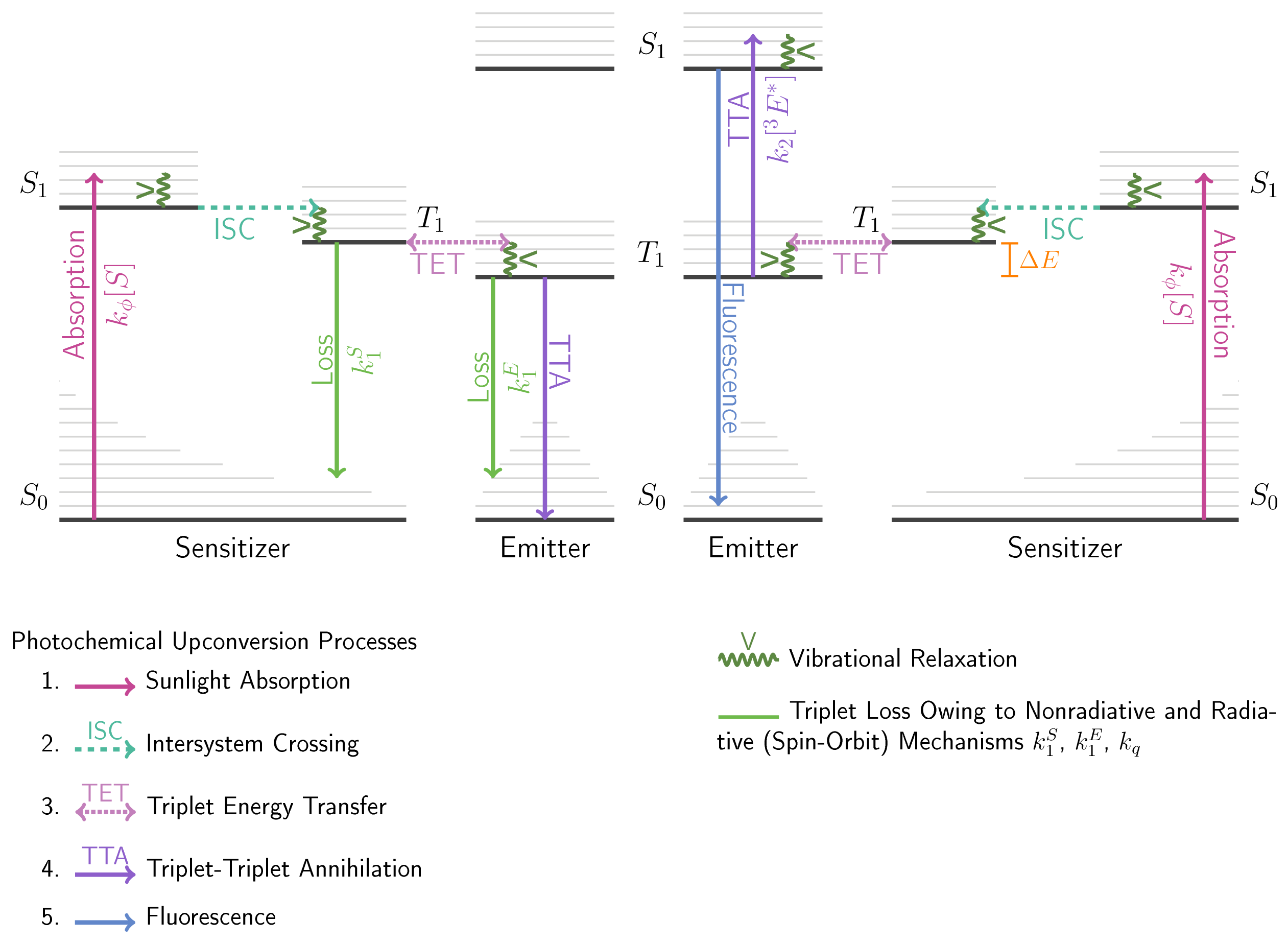}
    \caption{Photochemical upconversion energy level diagram.  The system consists of sensitizer and emitter molecules.  The sensitizer molecules capture light and transfer the resulting exciton to the emitter.  $S_n$ and $T_n$ indicate the $n$th singlet and triplet spin energy levels, respectively.  The triplet energy transfer double-ended arrows indicate that rapid triplet energy transfer achieves an equilibrium, rather than complete transfer. Figure adapted from \cite{gholizadeh2018photochemical}.}
    \label{fig:energylevel}
\end{figure*}

Photochemical upconversion transfers energy through a series of energy levels, which are illustrated by an energy level diagram in Fig. \ref{fig:energylevel}.  The energy levels are in two different molecules, the sensitizer \cite{wu2016solid,wu2015solid,okumura2016employing,amemori2015metallonaphthalocyanines,baluschev2007upconversion,huang2015hybrid} and the emitter \cite{turshatov2012synergetic,yu2015triplet,pun2018tips,gray2017loss,gaotetraphenylethene}.  First, sunlight is absorbed by the sensitizer molecules.  Second, the sensitizer undergoes intersystem crossing, which produces a triplet exciton.  Third, the sensitizer molecule transfers energy to an emitter molecule \cite{schmidt2014photochemical,cheng2016increased}. The triplet exciton state of the emitter is relatively long-lived \cite{gray2017loss,gholizadeh2018photochemical}, enabling energy to be stored for conversion. Fourth, emitter excitons undergo triplet annihilation.  Triplet annihilation converts a pair of triplet excitons to one singlet exciton.   Fifth, the emitter molecules in the singlet excited state produce fluorescence.  This fluorescence has a higher energy per photon than the light which was absorbed in the first step, so it can be used by a solar cell.  

Photochemical upconversion cannot exceed 50\% quantum yield because triplet annihilation converts two triplet excitons into one singlet exciton \cite{cheng2010kinetic}.  50\% quantum yield is highly advantageous because upconversion enables use of a region of the solar spectrum where the solar cell external quantum efficiency is zero \cite{tayebjee2015beyond}.  In addition, the output quanta have more energy than the input quanta; the energy efficiency exceeds the quantum yield.  

In upconversion devices, the balance between desirable triplet exciton annihilation and other forms of triplet loss determines the quantum yield of upconversion \cite{schmidt2014photochemical}.  Here, we use simulations \cite{frazer2017optimizing} to show how advanced triplet exciton physics can be applied to shift the fate of excitons towards annihilation and away from other decay mechanisms.  In particular, we focus on the Boltzmann distribution of triplets between molecules and the recently discovered quenching action of the sensitizer \cite{gholizadeh2018photochemical,schmidt2015photochemical} on the triplet exciton storage in the emitter.

\section{Overview of Calculations}

\begin{figure}
    \centering
    \includegraphics[width=\linewidth]{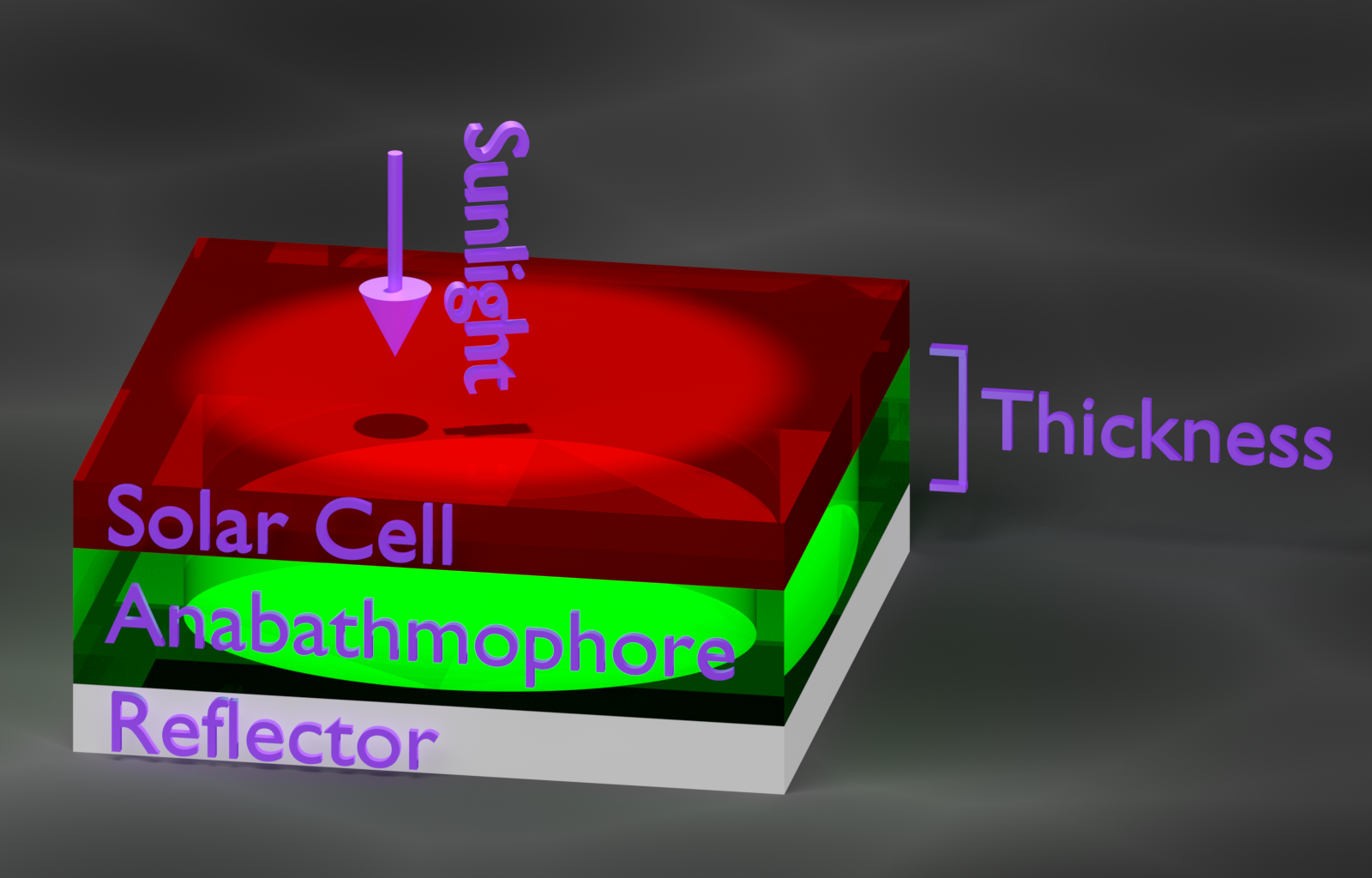}
    \caption{An illustration of the device, including the solar cell, light upconverting anabathmophore, and light distributing diffuse reflector.  The anabathmophore upconverts the light which is not absorbed by the solar cell owing to the bandgap.  The sensitizer and emitter are located in the anabathmophore.}
    \label{fig:device}
\end{figure}

We calculate the accepted figure of merit for photochemical upconversion, which is the photocurrent per unit area caused by upconversion, under the assumption that the solar cell has perfect quantum efficiency \cite{cheng2012improving}.  As illustrated in Fig. \ref{fig:device}, we simulate  a device consisting of a solar cell, an anabathmophore layer which performs photochemical upconversion \cite{frazerphotochemical}, and a Lambertian diffuse reflector \cite{frazer2017optimizing}.  Our simulations use random samples from the AM1.5G solar spectrum.  For each sunlight sample, we use random sampling to determine if the light is absorbed by the solar cell, absorbed by the sensitizer, or diffusely reflected.

It is established that, in well constructed systems, the sensitizer intersystem crossing, triplet energy transfer \cite{gray2017loss,monguzzi2008upconversion,huang2017pbs,younts2017efficient,nienhaus2017speed,gholizadeh2018photochemical,singh2009triplet}, and fluorescence have negligible losses.  Therefore, we assume they are perfectly efficient.  While our methods can be adapted to poorly constructed systems, including low intersystem crossing rates, triplet transfer rates, and fluorescence yields, these possibilities are beyond the scope of this report.  Triplet energy transfer is further discussed in Section \ref{sec:tet}.  

The quantum yield of photochemical upconversion $\Phi_\text{UC}$ was computed according to the accepted theory \cite{schmidt2014photochemical}, which incorporates the triplet exciton annihilation rate constant $k_2$, the triplet exciton concentration $[T]$, and the regular triplet loss rate constant $k_1$:
\begin{align}
    \Phi_\text{UC}&=\frac{k_2[T]}{2\left(k_1+k_2[T]\right)}\label{eq:yield}
\end{align}
This yield determines the quantity of fluorescence.

Using random samples from the fluorescence spectrum, we calculate the rate at which fluorescence is absorbed by the solar cell.  The figure of merit is calculated from this rate.   We also include the self-absorption and photon recycling \cite{pazos2016photon}, owing to both the sensitizer and the emitter, including diffuse reflections from the bottom surface of the anabathmophore.  Self-absorption is typically small for well-designed systems.

\begin{figure}
    \centering
    \includegraphics[width=.2\textwidth]{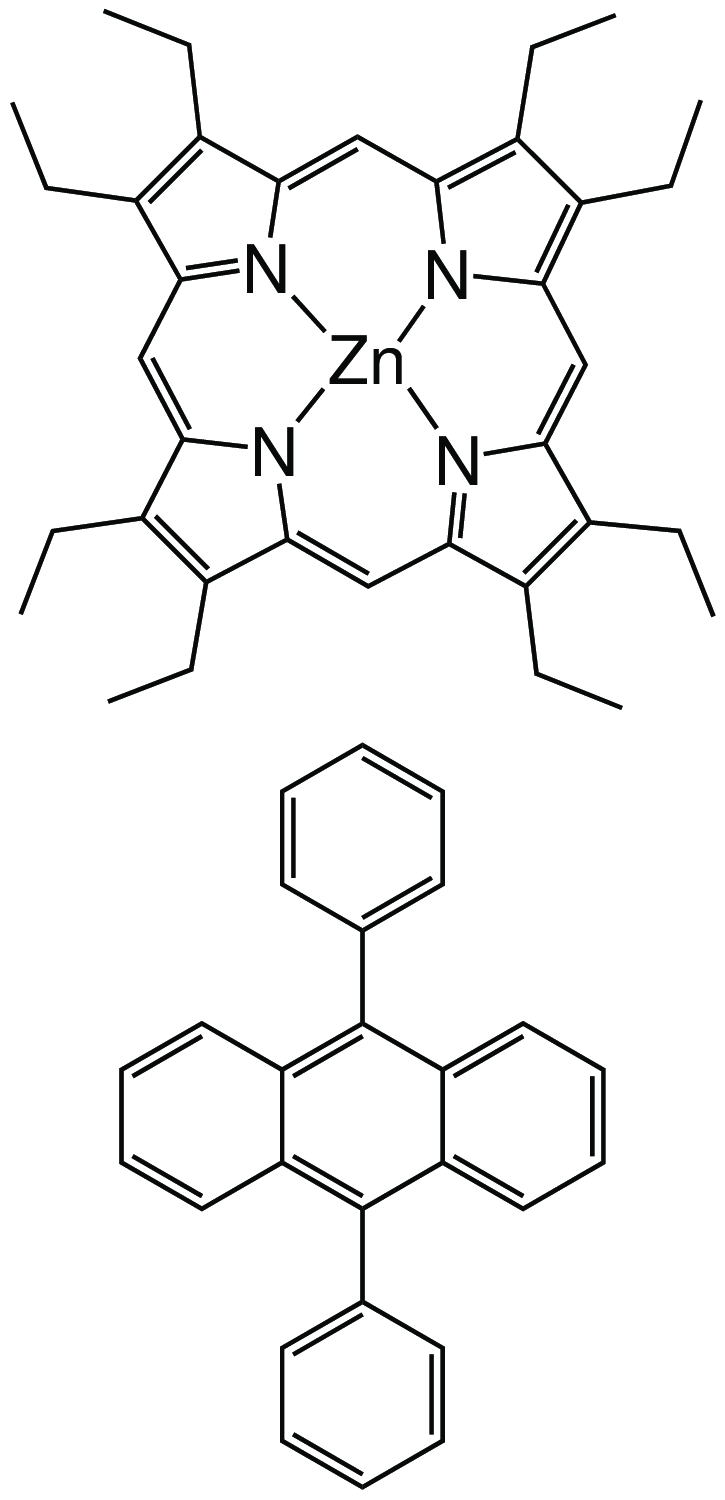}
    \caption{Molecular structures of sensitizer zinc octaethylporphyrin (top) and emitter 9,10-diphenylanthracene (bottom).}
    \label{fig:structures}
\end{figure}

\begin{figure}
    \centering
    \includegraphics[width=\linewidth]{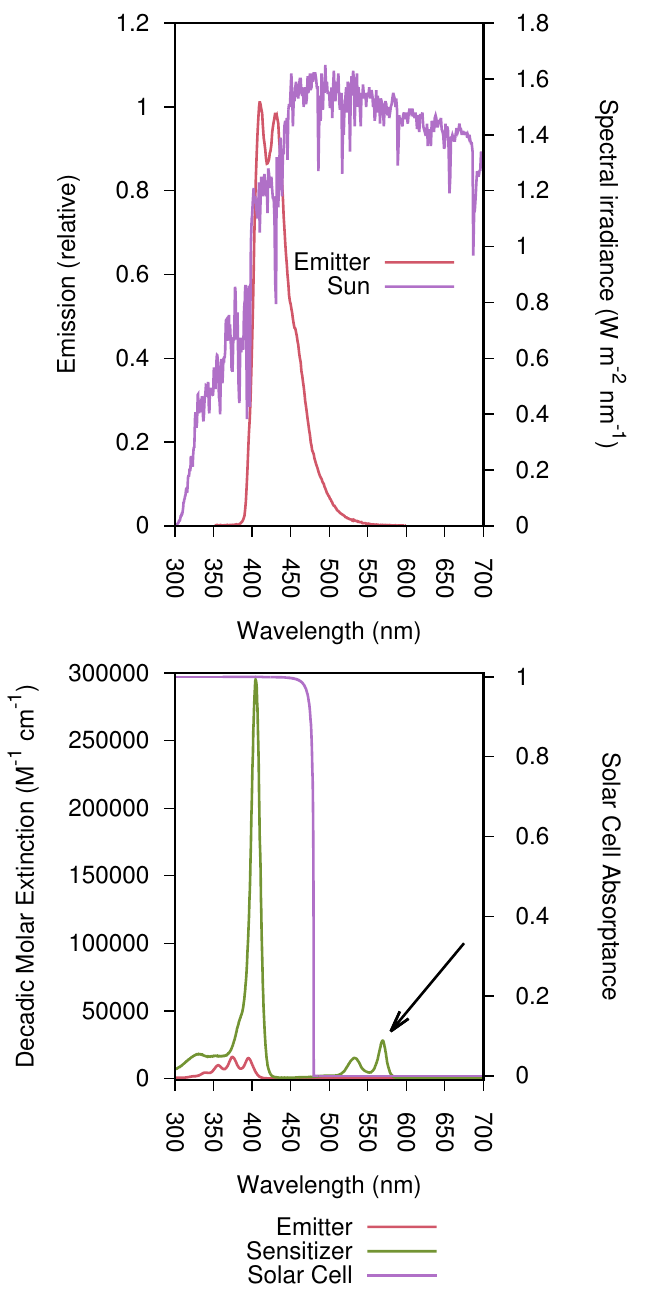}
    \caption{Example spectra, top:  Solar spectral irradiance and diphenylanthracene emitter fluorescence (normalized to peak).  Bottom: solar cell absorptance (with bandgap at 480\,nm), zinc octaethylporphyrin sensitizer molar extinction, and emitter molar extinction.  The arrow indicates the absorption peaks that sensitize upconversion.  Since the solar cell is simulated as a single interface, it has a dimensionless absorptance \cite{mcnaught1997compendium}.}
    \label{fig:spectra}
\end{figure}

Our simulations use experimental solar spectral irradiance, sensitizer absorption, emitter absorption, and emitter emission spectra.  Therefore, they are readily adapted to a wide range of illumination conditions and chemistries.  For this paper, we use the sensitizer zinc octaethylporphyrin and the emitter diphenylanthracene.  Previous work has investigated the relationship between the chemical structure and upconversion properties of closely related sensitizers \cite{aulin2015photochemical,gray2017loss,gholizadeh2018photochemical,turshatov2012synergetic,deng2015photochemical,wang2014efficient,baluschev2007upconversion,xun2016pd} and emitters \cite{gaotetraphenylethene,gray2017loss,gray2015photophysical,gray2016porphyrin,baluschev2008general,smith2010singlet}.  The chemical structures are illustrated in Fig. \ref{fig:structures} and the spectra are presented in Fig. \ref{fig:spectra}.  For the solar cell, we use the Tauc model of direct bandgap absorption so that the bandgap of the solar cell is a free variable \cite{viezbicke2015evaluation,tauc1968optical}.  Details of the algorithm are in Section \ref{sec:algorithm}.

\section{Triplet kinetics owing to triplet energy levels}\label{sec:tet}
The rate of energy transfer from the sensitizer to the emitter is much faster than the triplet decay rate of the sensitizer \cite{gray2017loss,monguzzi2008upconversion,huang2017pbs,younts2017efficient,nienhaus2017speed,gholizadeh2018photochemical,singh2009triplet}.  As a result, it is common to discuss the decay rate of triplet excitons (excluding annihilation) $k_1$ as if it were the same as the triplet decay rate of the emitter.  However, we will show that the sensitizer triplet decay rate can play an important role in determining the figure of merit, even though energy transfer is faster than triplet decay.

In equilibrium, the distribution of triplets between the sensitizer and the emitter is according to the Boltzmann distribution \cite{cheng2011entropically}.  Experiments show the triplet energy transfer is the fastest rate when the emitter concentration is high \cite{gray2017loss}. We assume the sensitizer and emitter triplet exciton populations are in equilibrium.  The equilibrium is illustrated in Fig. \ref{fig:energylevel} by the triplet energy transfer arrows from the sensitizer to the emitter and from the emitter to the sensitizer.

If $k_1^S$ is the triplet decay rate constant in the sensitizer \cite{gray2017loss}, $k_1^E$ is the triplet decay rate constant in the emitter \cite{gholizadeh2018photochemical}, $[S]$ is the sensitizer concentration, $[E]$ is the emitter concentration, $k_B$ is the Boltzmann constant, $T$ is the temperature, and $\Delta E$ is the difference between the sensitizer triplet energy level and the emitter triplet energy level, then the overall triplet decay rate is  
\begin{align}
k_1&=\frac{[S]k_1^S e^{-\frac{\Delta E}{k_B T}}+[E]k_1^E}{[S]e^{-\frac{\Delta E}{k_B T}}+[E]}.\label{eq:k1boltz}
\end{align}
The rate constants are listed in Table \ref{tab:rates} \cite{gray2017loss,gholizadeh2018photochemical}.
\begin{table*}
    \centering
    \caption{Physical rate constants assumed in the simulation.}
    \begin{ruledtabular}
    \begin{tabular}{lclrc}
    Name&Symbol&Compound&Value&Reference\\\hline
    \rule{0pt}{3ex}Triplet decay of sensitizer&$k_1^S$&Zinc Octaethylporphyrin&\num{8550}\,\si{\per\s}&\cite{gray2017loss}  \\
    Triplet decay of emitter, $[S]=0$&$k_1^0$&Diphenylanthracene&\num{2000}\,\si{\per\s}&\cite{gholizadeh2018photochemical}  \\
    Quenching&$k_q$&Both&\num{4.8e7}\,\si{\per\Molar\per\s}&\cite{gholizadeh2018photochemical}\\
    &&&\SI{8.0e-14}{\cm\cubed\per\s}\\
    Triplet annihilation&$k_2$&Diphenylanthracene&\SI{2.8e9}{\per\Molar\per\s}&\cite{gray2017loss,gray2015photophysical}\\
	 &&&\num{4.7e-12}\,\si{\cm\cubed\per\s}
    \end{tabular}
    \label{tab:rates}
    \end{ruledtabular}
\end{table*}

\begin{figure}
    \centering
    \includegraphics[width=\linewidth]{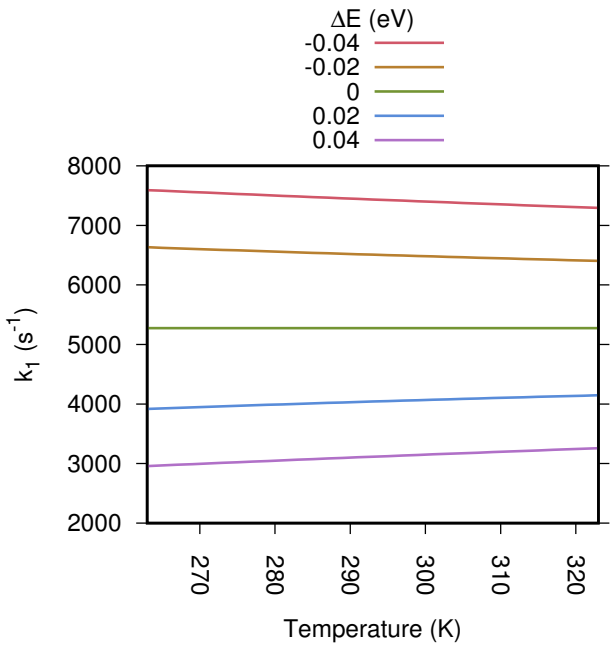}
    \caption{$k_1$ as a function of temperature for several values of $\Delta E$ according to the statistical distribution of triplet excitons.   Smaller $k_1$ leads to a better $\Phi_\text{UC}$ (Equation \ref{eq:yield}).  Better upconversion systems are exothermic ($\Delta E\gg k_BT$), so their $k_1$ increases when heated by sunlight.  Despite the detrimental effect of heating, exothermic systems remain superior.  For this figure, we use $[S]=[E]=$\,1\,\si{\milli\Molar}.}
    \label{fig:exothermic}
\end{figure}

As shown in Fig. \ref{fig:exothermic}, if $\Delta E>0$, then as temperature increases, $k_1$ increases, which leads to reduced quantum yield.  This result emphasizes that the interplay between environmental conditions, such as heating by the sun and cooling by the wind \cite{skoplaki2009temperature}, with the exothermic nature of upconversion must be considered when designing an energy conversion system.

In order to generate triplet excitons, the sensitizer must have a high intersystem crossing rate.  As a result, the sensitizer's triplet decay rate constant $k_1^S$ is relatively large \cite{perun2008singlet,mongin2016direct,gray2017loss}.  However, device designers have the freedom to select an emitter molecule with a small triplet decay rate constant $k_1^E$.  Equations \ref{eq:yield} and \ref{eq:k1boltz} show that a large $k_1^S$ decreases the quantum yield.  

In addition, a pair of triplet states located in the sensitizer typically cannot produce upconversion because known sensitizer molecules lack a suitable first excited singlet spin state \cite{kasha1950characterization}.   It is possible to harvest higher excited states \cite{goudarzi2019impact}, a phenomenon which we do not simulate.  While annihilation of triplet excitons located in different molecules can be efficient \cite{cao2013high}, we assume that sensitizers will not have this property.  Therefore, the concentration of \emph{usable} triplet excitons is
\begin{align}
    [^3E^*]&=[T]\frac{[E]}{[S]e^{-\frac{\Delta E}{k_BT}}+[E]}.
\end{align}
$\Delta E$ is the energy lost during transfer of a triplet exciton from the sensitizer to the emitter.  It is indicated in Fig. \ref{fig:energylevel}.
The quantum yield is more precisely written as
\begin{align}
    \Phi_\text{UC}&=\frac{k_2[^3E^*]}{2\left(k_1+k_2[^3E^*]\right)}.\label{eq:yield2}
\end{align}

In many cases, $\Delta E$ much is larger than the thermal energy.  Then it is not necessary to consider $k_1^S$. An excessively large $\Delta E$ is detrimental to energy efficiency because it reduces the upward shift in the energy of the upconverted photons.  If $\Delta E$ is enhanced by shifting the sensitizer triplet energy upwards, then the portion of the solar spectrum which could be captured by the sensitizer will decrease.  If $\Delta E$ is enhanced by shifting the emitter triplet energy downwards, it may be necessary to also shift the fluorescence energy downwards to keep the system exothermic.  This leads to the necessity of selecting a solar cell with a smaller bandgap.  The solar cell will then block more light from reaching the sensitizer.  

If $\Delta E$ is negative, it is possible to increase the photon energy by more than a factor of two  \cite{cheng2011entropically}.  However, more triplet excitons will be distributed in the sensitizer molecules, which makes the quantum yield small.

\begin{figure}
    \centering
    \includegraphics[width=\linewidth]{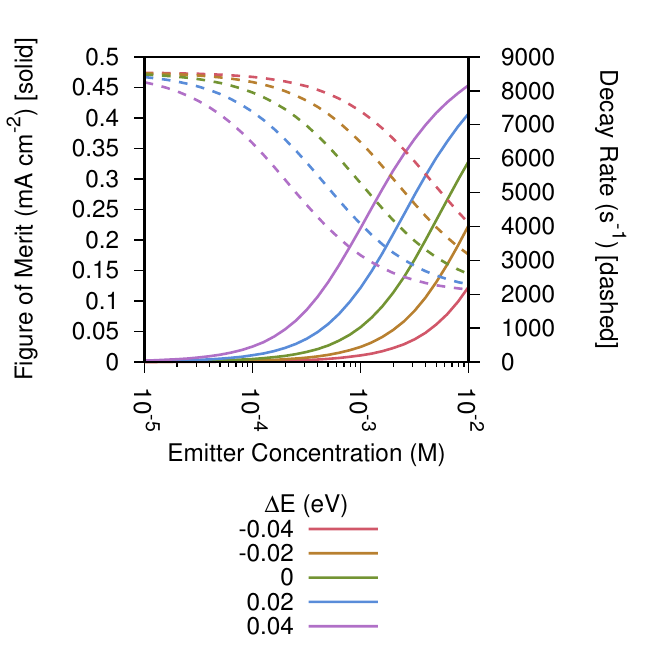}
    \caption{Triplet decay rate constant $k_1$ (dashed curves) and figure of merit (solid curves) as a function of emitter concentration $[E]$ for several values of $\Delta E$.  The sensitizer concentration is 1\,\si{\milli\Molar}, temperature is \SI{300}{\K} and anabathmophore thickness is \SI{0.1}{\cm}.  If $\Delta E$ or $[E]$ are not big enough, upconversion becomes inefficient because triplets decay in the sensitizer.  A large triplet energy transfer rate cannot overcome this decay.  In addition, triplet excitons in the sensitizer are not available for upconversion.}
    \label{fig:k1concentration}
\end{figure}

Fig. \ref{fig:k1concentration} shows the simulated figure of merit, current density, as a function of emitter concentration.  For low values of $\Delta E$, a high ratio of emitter molecules to sensitizer molecules is needed to produce upconversion.  An abundance of emitter molecules ensures some triplet excitons are distributed to the emitter.  If $\Delta E$ is large, then the Boltzmann distribution ensures triplet excitons are located in the emitter molecules even if those molecules are scarce.  We do not include the additional decrease in the figure of merit which occurs at very low emitter concentrations because the triplet excitons do not reach the equilibrium distribution before decaying, or the emitter becomes saturated with excitations.

\begin{figure}
    \centering
    \includegraphics[width=\linewidth]{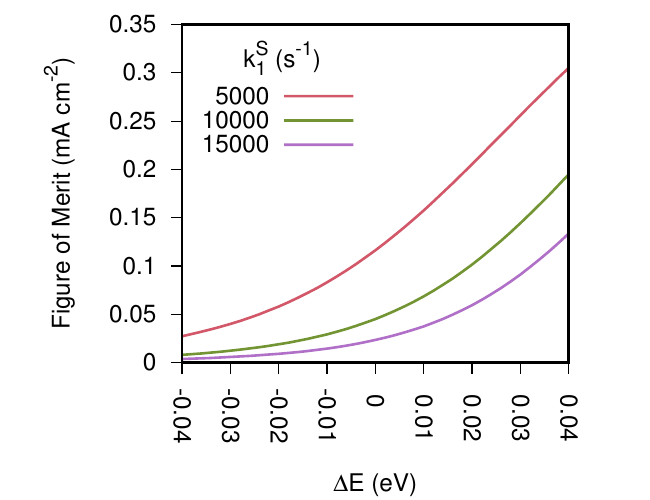}
    \caption{Figure of merit as a function of difference in triplet energy levels $\Delta E$, for a variety of sensitizer triplet decay rate constants $k_1^S$ and anabathmophore thickness of \SI{0.1}{\cm}.  Energy losses owing to $k_1^S$ are mitigated if $\Delta E\gg k_BT$.  Platinum octaethylporphyrin with diphenylanthracene exhibits this advantage.  Zinc octaethylporphyrin does not.  However, the spectral shift achieved by upconversion decreases as $\Delta E$ increases.  For this figure, we use $[S]=[E]=$\,1\,\si{\milli\Molar}.}
    \label{fig:exothermicE}
\end{figure}

Fig. \ref{fig:exothermicE} shows that upconversion fails to produce photocurrent when all the triplet decay occurs in the sensitizer because $\Delta E\ll 0$.  
Smaller sensitizer triplet decay $k_1^S$ makes the figure of merit more sensitive to $\Delta E$ near $\Delta E=0$. The zinc octaethylporphyrin/diphenylanthracene system is in this region.  Both $\Delta E$ and $k_1^S$ can contribute to controlling triplet exciton loss.

\section{Dynamic triplet quenching caused by sensitizer}
Sensitizer concentration determines the excitation density \cite{swinehart1962beer}.  A high excitation density produces efficient annihilation because the triplet concentration is in the numerator of Equation \ref{eq:yield2}.  Therefore, one would expect that the highest achievable sensitizer concentration will produce the highest possible upconversion figure of merit.  We have recently shown that $k_1^E$, the triplet decay in the emitter, is dependent on sensitizer concentration \cite{gholizadeh2018photochemical}.  Here we show the resulting impact on the figure of merit and device design.

\subsection{Model}
The decay rate of triplets in the emitter in the absence of sensitizer and excluding annihilation is $k_1^0$, which typically ranges from $10^2$ to $10^4$\,\si{\per\second} \cite{gholizadeh2018photochemical}.  The rate constant quantifying emitter triplet quenching by the sensitizer, $k_q$, is typically less than \num{5e7}\,\si{\per\Molar\per\second} \cite{gholizadeh2018photochemical}. Total triplet losses in the emitter are
\begin{align}
k_1^E&=k_1^0+k_q[S].\label{eq:quench}
\end{align}
The solubility limit on $[S]$ is above \SI{1}{m\Molar} \cite{gholizadeh2018photochemical}.

\subsection{Quenching reduces figure of merit}
\begin{figure}
    \centering
    \includegraphics[width=\linewidth]{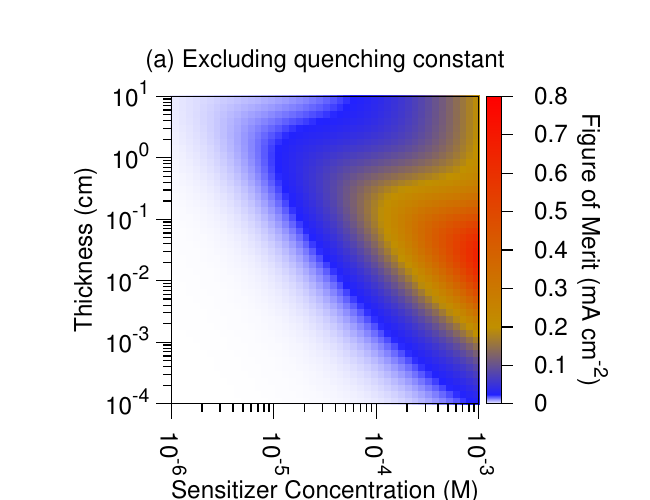}
    
    \includegraphics[width=\linewidth]{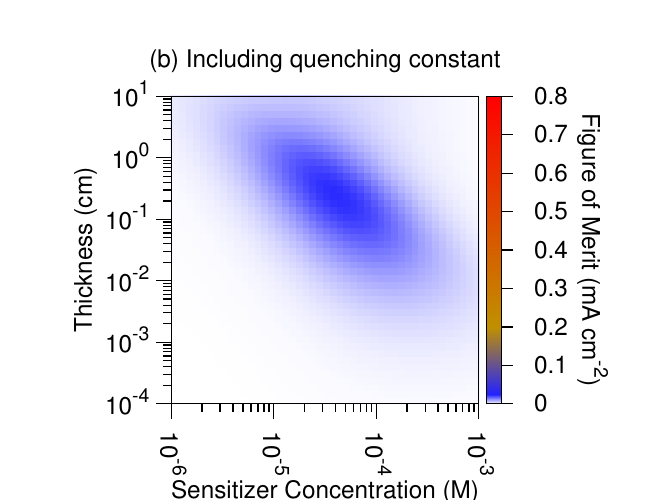}
        \caption{Figure of merit as a function of sensitizer concentration and thickness, (a) without  and (b) with the quenching constant $k_q=$ \num{4.8e7}\,\si{\per\Molar\per\s} \cite{gholizadeh2018photochemical}.  White indicates zero figure of merit.  We assume $\Delta E=$\,\num{0.3}\,eV.}
    \label{fig:concentration}
\end{figure}

Fig. \ref{fig:concentration} shows the calculated figure of merit with and without the quenching constant.   Here, we assume strongly exothermic triplet energy transfer.  The device thickness, which is important to achieving a high triplet concentration \cite{frazer2017optimizing}, is also included as a variable.  Without the quenching constant, the figure of merit increases with concentration. The optimal device thickness decreases with concentration, as the absorption length decreases.  With the inclusion of a quenching constant, a maximum figure of merit exists below \num{e-4}\,\si{\Molar} sensitizer concentration.

\begin{figure}
    \centering
    \includegraphics[width=\linewidth]{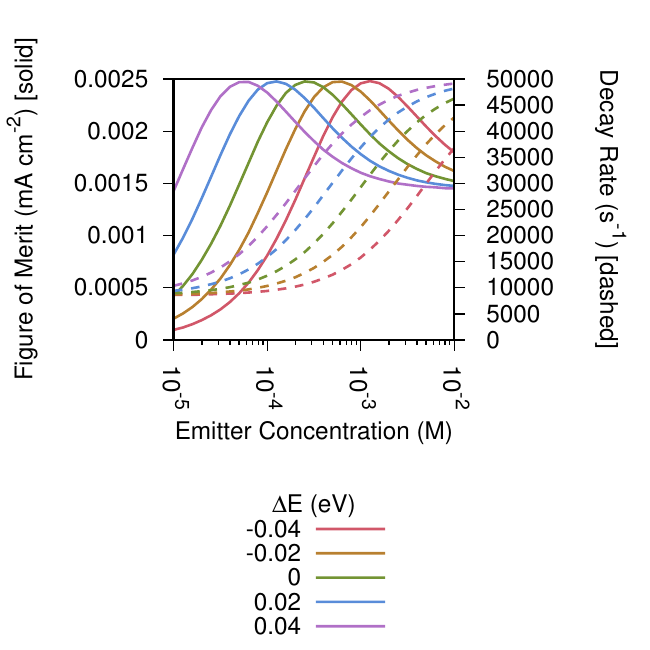}
    \caption{Triplet decay rate constant $k_1$ (dashed curves) and figure of merit (solid curves) as a function of emitter concentration for several values of $\Delta E$ with the quenching constant of \num{4.8e7}\,\si{\per\Molar\per\s}.  The sensitizer concentration is 1\,\si{\milli\Molar}, temperature of \SI{300}{\K} and thickness is $\SI{0.1}{\cm}$.  Compared to Fig. \ref{fig:k1concentration}, the figure of merit is reduced by the quenching action of the sensitizer.  The figure of merit has a maximum for the same reason.}
    \label{fig:k1concentrationwithkq}
\end{figure}

The relationship between the figure of merit and the emitter concentration is dramatically changed by the inclusion of the quenching constant.  In our model, the triplet excitons in the sensitizer are protected from concentration-dependent quenching.  As a result, in Fig. \ref{fig:k1concentrationwithkq}, the triplet decay rate $k_1$ increases with emitter concentration.  Unlike the results in Fig. \ref{fig:k1concentration}, which exclude quenching, the figure of merit has a maximum with respect to emitter concentration when quenching is included.  We omit the concentration quenching of triplet excitons within triplet sensitizers, which can be $\sim$\SI{e7}{\per\Molar\per\second}, and may be important \cite{callis1973porphyrins,dexter1954theory}.

\subsection{Interplay of sensitizer quenching and Boltzmann statistics}
The quenching effect of Equation \ref{eq:quench} was experimentally demonstrated in situations where $\Delta E\gg k_BT$ \cite{gholizadeh2018photochemical}.  As the sensitizer concentration increased, $k_1$ increased in a linear fashion.  If this analysis is performed on a system that is not strongly exothermic, then the contribution of Boltzmann statistics from Equation \ref{eq:k1boltz} will cause the quenching constant to be overestimated.  For zinc octaethylporphyrin and diphenylanthracene, the reported $\Delta E$ is 0.02\,eV \cite{gray2017loss,ohno1985luminescence,wu1991photophysical,liu2006photophysics,brinen1968lowest}.  In our view, the experimental and theoretical uncertainty on this value is enough that the sign is uncertain.   

\begin{figure}
    \centering
    \includegraphics[width=\linewidth]{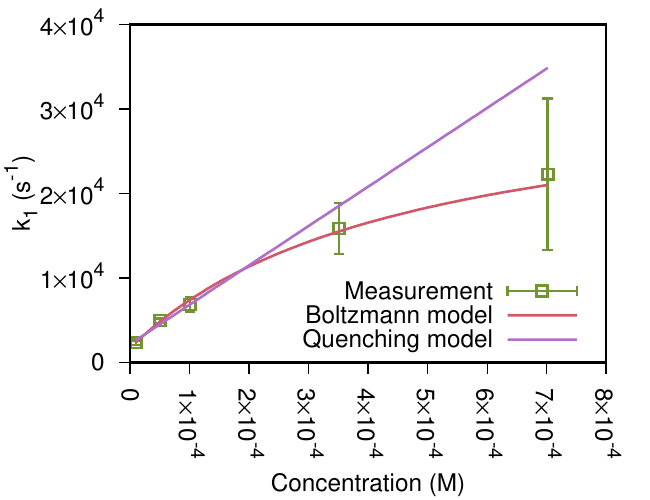}
    \caption{Experimental triplet decay rate in the emitter as a function of sensitizer concentration. The Linear Quenching Model \ref{eq:quench} is compared against the Boltzmann Model including Equations \ref{eq:k1boltz} and \ref{eq:quench}.  The downward curvature of the data suggests some triplet excitons remain in the sensitizer after triplet energy transfer reaches equilibrium.  Data from \cite{gholizadeh2018photochemical}.}
    \label{fig:quenchreanalysis}
\end{figure}

In Fig. \ref{fig:quenchreanalysis}, we reanalyze zinc octaethylporphyrin and diphenylanthracene data from Ref. \cite{gholizadeh2018photochemical}.  We compare the prediction of Equation \ref{eq:quench} with the combined prediction of Equations \ref{eq:k1boltz} and \ref{eq:quench}.  Equation \ref{eq:k1boltz} increases the number of free parameters, so its inclusion must improve the accuracy of the model.  While the experimental uncertainty is large enough that neither model can be rejected, it does seem that $\Delta E$ is not large enough to keep all the triplet excitons in the emitter.  We suggest that $\Delta E =$\,\SI{-0.02\pm0.01}{eV}.  The triplet energy transfer may be endothermic.

If the mechanism giving rise to $k_q$ were an external heavy atom effect, then it should increase with atomic number $Z$.  However, the opposite was observed \cite{gholizadeh2018photochemical}.  Zinc-containing sensitizer ($Z=30$) had the highest $k_q$, but palladium ($Z=46$) and platinum ($Z=78$) were similar to each other.  Inclusion of $\Delta E$ in the theory opens up the possibility that zinc-containing sensitizer does not really have a higher $k_q$. Both parameters can explain the experimental increase in the emitter triplet exciton decay. Future measurements over a range of emitter concentrations will eliminate this ambiguity.   $\Delta E$ and $k_q$ provide theoretical explanations of the relationship between atomic number and upconversion performance that do not require models based on phenyl group rotation \cite{gray2017loss}. Table \ref{tab:relativeimportance} shows that both $\Delta E$ and $k_q$ can change the figure of merit, but that $k_q$ is more important.
\begin{table}
    \centering
    \begin{ruledtabular}
    \begin{tabular}{rrr}
    $\Delta E$\,(\si{\eV})&$k_q$\,(\si{\per\Molar\per\second})&$J_\text{UC}$\,(\si{\mu\A\per\cm\squared})\\\hline
    \rule{0pt}{3ex}\num{-.02}&\num{4.8e7}&\num{23.5\pm0.1}\\
    \num{-.02}&\num{0}&\num{370.0\pm0.5}\\
    \num{0.3}&\num{4.8e7}&\num{24.2\pm0.1}\\
    \num{0.3}&\num{0}&\num{673.2\pm0.1} \\
    \end{tabular}
    \caption{Figure of merit $J_\text{UC}$ under different assumptions about $\Delta E$ and $k_q$.  The sensitizer concentration and device thickness are optimized separately for each calculation.  With no $k_q$ the sensitizer concentration is constrained to 1\,\si{m\Molar} by solubility.}
    \label{tab:relativeimportance}
    \end{ruledtabular}

\end{table}

\subsection{Selecting the optimal design}
\begin{figure}
    \centering
    \includegraphics[width=\linewidth]{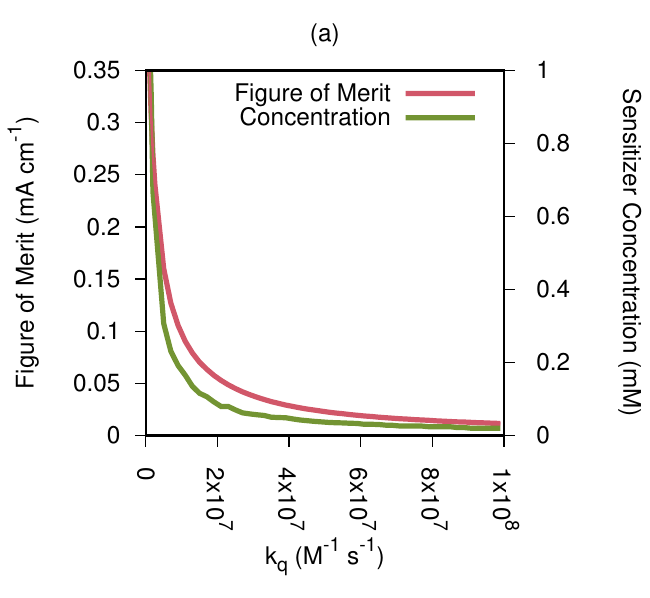}
    \includegraphics[width=\linewidth]{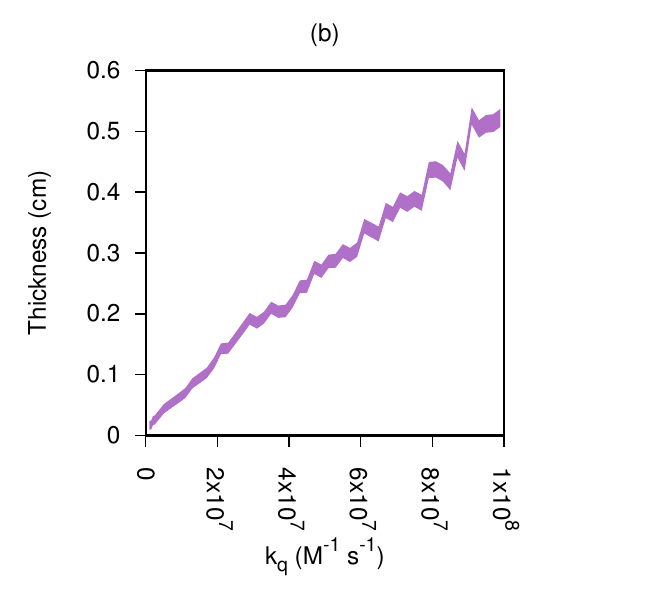}
    \caption{(a) Maximum figure of merit and optimal sensitizer concentration as a function of quenching constant. (b) Optimal thickness as a function of quenching constant.  We assume $\Delta E=0$ and $[E]=$\,\num{e-2}\,\si{\Molar}.  The width of the purple curve indicates estimated Monte Carlo error.}
    
    \label{fig:kq}
\end{figure}

Fig. \ref{fig:kq} gives the optimized figure of merit as a function of quenching constant.  This shows the harmful effect of the sensitizer quenching the emitter on device performance.  In addition, $k_q$ makes it necessary to increase the device thickness and decrease the sensitizer concentration to generate the most current from an upconversion device.  The quantity of sensitizer used may be important to cost effectiveness.

\section{Figure of Merit Algorithm\label{sec:algorithm}}
\subsection{Sampling Sunlight}
The device was modeled %American English please
as an infinite plane with the sunlight incident perpendicular to the surface. The solar spectral irradiance was stochastically sampled $10^9$ times using the cumulative distribution function of the AM1.5G spectrum. To determine if a sample was absorbed into the solar cell, the Tauc model of direct bandgap semiconductors \cite{tauc1968optical,viezbicke2015evaluation} was scaled so there was a \num{99}\% probability of the solar cell absorbing the sample \num{0.1}\,eV above the 2.6\,\si{eV} bandgap. If the sample was not absorbed, it was assumed to reach the anabathmophore.   Refractive index matching was assumed throughout the simulation.
\subsection{Light Transmission and Scattering}

The interior of the anabathmophore was divided into $10^5$ bins arranged vertically. Bins were used to model the inhomogeneous distribution of excitons within the anabathmophore. 

If a sample reached the anabathmophore, spline interpolation was used to calculate the absolute value of the corresponding sensitizer and emitter molar absorptivities from the experimental spectra shown in Fig. \ref{fig:spectra}. The emitter absorption spectrum was filtered so any molar extinction below 1000\,\si{\per\Molar\per\cm} was set to 0\,\si{\per\Molar\per\cm} to mitigate instrument noise.  This reduces nonphysical anti-Stokes shifts.  Using the sum of the molar absorptivities, the distance travelled by the sample was stochastically determined from the Beer-Lambert Law \cite{swinehart1962beer}.  The concentrations and molar absorptivities were used to stochastically assign the sample to be absorbed by the sensitizer or emitter. From the distance travelled, the bin the sample was absorbed into was determined.  For each bin, the number of samples absorbed and reabsorbed by the sensitizer and the emitter were recorded.

If the distance travelled exceeded the thickness of the anabathmophore, Lambertian reflection was simulated and the distance travelled was recalculated. The upward component of the distance travelled was used to determine the bin.   If the distance travelled upward went beyond the region occupied by the anabathmophore, the sample was not absorbed by the anabathmophore and did not contribute to the figure of merit.

\subsection{Sampling Fluorescence}
Samples absorbed by the sensitizer and emitter were modelled separately and had different fluorescence yields. 
For sensitizer excitation, we assumed the singlet yield of triplet annihilation was perfect \cite{hoseinkhani2015achieving,cheng2010kinetic}.  
The quantum yield $\Phi_\text{UC}$ was calculated using Equation (\ref{eq:yield2}) with values from Table \ref{tab:rates}.  Temperature was assumed to be \SI{300}{\K}. Since triplet energy transfer is rapid \cite{gray2017loss}, triplet energy transfer was assumed to be in equilibrium.  The triplet concentration was calculated using \cite{schmidt2014photochemical}
\begin{align}
  [T]&=\frac{-k_1+\sqrt{k_1^2+4k_\phi k_2[S]}}{2k_2},
\end{align} where $k_\phi$ is the excitation rate computed under the usual assumption that the irradiance of the sun is \SI{1}{\kilo\W\per\m\squared}. The upconversion yield $\Phi_\text{UC}$ was different for each bin because $k_\phi$ was different.  Samples absorbed by the emitter had perfect fluorescence yield, corresponding to an assumption of perfect fluorescence quantum yield.  The quantum yield assumption also applied to converted triplet excitons.

To simulate emission, fluorescence samples generated according to the quantum yield were each stochastically assigned a wavelength.   The wavelength was determined using the cumulative distribution function of the experimental emitter fluorescence spectrum in Fig. \ref{fig:spectra}. The fluorescence was propagated in a random direction from the middle of the bin according to the Beer-Lambert Law. If the direction of travel was downward, the sample could undergo Lambertian reflection. The vertical component of distance travelled determined the bin where the sample was reabsorbed. If the sample escaped from the top of the anabathmophore, the Tauc model was used again to determine if the solar cell absorbed the sample \cite{viezbicke2015evaluation,tauc1968optical}.  If it did, then the sample contributed to the figure of merit. 

Reabsorption and emission were recalculated five times to account for photon recycling \cite{pazos2016photon}.  During each cycle, the number of samples reabsorbed in each bin and $\Phi_\text{UC}$ were recalculated.  Typically, reabsorption was small.  Here, we chose a sensitizer which was mostly transparent to fluorescence; our previous results suggest that molar extinction is more important to the figure of merit than reduced reabsorption, so it is important to account for photon recycling \cite{frazer2017optimizing}.

\subsection{Figure of Merit}
The total radiant exposure entering the system was calculated by summing the photon energies of samples stochastically generated from the solar spectrum. The radiant exposure was then divided by the standard solar irradiance, \SI{1}{\kilo\W\per\m\squared}, to find the simulation duration $t$.  The figure of merit, current density, was
\begin{align}
    \frac{en}{t}
\end{align}
where $e$ is the fundamental charge and $n$ is the area density of emitter fluorescence samples absorbed by the solar cell.
\section{Conclusions}
Previously, we argued that \SI{0.1}{\milli\ampere\per\cm\squared} is a meaningful figure of merit \cite{frazer2017optimizing}.  Reaching this goal requires improvement.  The efficiency of photochemical upconversion relies on the exothermic nature of each process involved in the steps of light conversion.  Here, we relate a decrease in the potential energy which drives triplet energy transfer to a reduction in the figure of merit.  We quantitatively demonstrated that the photocurrent is improved when  the energy loss $\Delta E\gg k_B T$.  The quenching of triplets located in the emitter but caused by the sensitizer also inhibits the figure of merit, even when exothermic operation is successfully achieved.  In the future, upconversion can be advanced by creating sensitizer/emitter pairs which have a large $\Delta E$ and a small quenching $k_q$.

The poor performance of zinc octaethylporphyrin with diphenylanthracene, compared to alternate sensitizers paired with diphenylanthracene, can be explained by the quenching action of the sensitizer on the emitter and the alignment of triplet energy levels.  These properties cannot be determined by only measuring triplet energy transfer rate constants.  The Boltzmann distribution of triplet excitons inhibits upconversion even when energy transfer is efficient.  

Our results highlight the value of measuring triplet decay rates as a function of concentrations.  This type of experiment provides information about exciton conversion processes which lead to major changes in the figure of merit.  
When $\Delta E\approx 0$, theoretical and phosphorescence methods may not be precise enough. In addition, a pulsed experiment showing a high triplet energy transfer rate is not sufficient to show complete triplet transfer in equilibrium.  If the Boltzmann factor is in doubt, then both the sensitizer and emitter concentration should be explored to find $\Delta E$ and $k_q$.

The simulation program, which is available from \mbox{\url{http://laszlofrazer.com}}, can readily be used to predict the relative merits of different compounds with respect to their usefulness as sensitizers and emitters.  This is particularly useful for the search for sensitizers which have absorption spectra that efficiently capture sunlight.  In addition, different solar cell bandgaps, illumination conditions, and rate constants can be conveniently modeled.  The manual is included as supplemental material \footnote{See Supplemental Material at [URL will be inserted by
publisher] for the manual for the upconversion solar cell simulator.}.

We have shown that, using knowledge of the sensitizer and emitter physics, the energy conversion performance of complete devices can be simulated.  These simulations enable prediction of the best device design without requiring construction of many devices with different molecular concentrations and device geometries.

\begin{acknowledgments}
This work was supported by the Australian Research Council Centre of Excellence in Exciton Science (CE170100026).  D. J. acknowledges a ResearchFirst fellowship from Monash University.  This research was undertaken with the assistance of resources and services from the National Computational Infrastructure (NCI), which is supported by the Australian Government.  We thank Elham Gholizadeh and Rowan MacQueen for experimental spectra used in this work.
\end{acknowledgments}
\bibliography{bibliography}
\end{document}